\documentclass[12pt]{article}
\usepackage[margin=72.27pt]{geometry}
\baselineskip=\the\baselineskip plus 0.2pt minus 0.2pt\relax
\usepackage[verbose=silent,factor=700,stretch=14,shrink=14,step=1]{microtype}
\usepackage{tikz}
\usetikzlibrary{cd}
\PassOptionsToPackage{implicit=false}{hyperref}
\usepackage{orcidlink}
\usepackage[font=small,labelfont=small,bf]{caption}
\usepackage{enumitem}
\usepackage{amsmath}
\usepackage{amssymb}
\usepackage{amsthm}
\DeclareMathOperator{\BR}{BR}
\DeclareMathOperator{\id}{id}
\DeclareMathOperator{\tor}{Tor}
\newtheorem{lemma}{Lemma}
\newtheorem{theorem}[lemma]{Theorem}
\newtheorem{corollary}[lemma]{Corollary}
\usepackage[authordate,doi=false,footmarkoff]{biblatex-chicago}
\def\enr{\textsc{enr}}

\begin{document}

{\LARGE\topskip=0.7in
\centerline{Nash Equilibrium Existence without Convexity}\smallskip
\centerline{Conrad Kosowsky\kern0.1em\relax\orcidlink{0009-0007-0479-6746}%
  \def\thefootnote{*}%
  \footnote{University of Michigan, Department of Economics. Email: coko@umich.edu.}%
  \addtocounter{footnote}{-1}}
}

\vskip 0.3in

\begin{abstract}
In this paper, I prove the existence of a pure-strategy Nash equilibrium for a large class of games with nonconvex strategy spaces. Specifically, if each player's strategies form a compact, connected Euclidean neighborhood retract and if all best-response correspondences are null-homotopic, then the game has a pure-strategy Nash equilibrium. As an application, I show how this result can prove the fundamental theorem of algebra.

\medskip


\medskip

\end{abstract}

\vskip 0.1in

Nash equilibrium is a central concept in game theory and has found applications throughout social science. Existence proofs typically rely on the Kakutani fixed-point theorem or the more general Fan-Glicksberg fixed-point theorem, and these theorems depend crucially on certain convexity assumptions.\footnote{See \textcite{tanaka-2011,tanaka-2012} for a constructive proof of these two results and an analysis of their relationship to the Brower fixed-point theorem.} The usual approach is to apply a fixed-point theorem to the best-response correspondence, so many standard techniques fail to establish Nash equilibrium for games without convex strategy spaces or best responses. In this paper, I apply Nielsen theory, which is a technique from algebraic topology, to prove equilibrium existence for a much broader class of games where strategy spaces and best responses may be nonconvex. I show that if each player's strategies form a compact, connected Euclidean neighborhood retract and if all best-response correspondences are null-homotopic, then the game has a pure-strategy Nash equilibrium, and I discuss topological concerns that can prevent equilibrium otherwise. This is a new and general result establishing the existence of pure-strategy Nash equilibrium without convexity requirements, and I show how to use it to prove the fundamental theorem of algebra.

Previous work on Nash equilibrium existence has almost entirely focused on games where all strategy spaces and best-response sets are convex, although a small recent literature discusses nonconvex games, typically from an algorithmic perspective. \textcite{pang-scutari-2011}, \textcite{scutari-pang-2013}, and \textcite{siyari-etal-2017} develop a related solution concept called quasi-Nash equilibrium for nonconvex games, and more recent work by \textcite{liu-etal-2021} focuses on approximation methods of $\epsilon$-Nash equilibrium. Games with nonconvexities arise for example in machine learning where finding a minmax solution to a smooth function that may not be convex or concave is important. See \textcite{nouiehed-etal-2019} and \textcite{ostrovskii-etal-2021} for recent work on this topic. In particular, \textcite{liu-etal-2021} write that ``when the players' cost functions are nonconvex and/or their action sets are nonconvex but not necessarily finite, there is no general result for the existence of [pure-strategy Nash equilibria].'' My paper addresses this gap in the literature.

The paper is organized as follows. Section~1 discusses the mathematical framework for the paper, difficulties that arise from arbitrary strategy sets, and the results we will use from Nielsen theory. Sections~2 and 3 prove existence for an example case and the general case respectively. Section~4 contains an example application, and Section~5 concludes.

\section{Mathematical Framework}

Formally, a \textit{game} is a collection of topological spaces $(X_i)$ indexed by $i\in I$ where $I$ is some index set. We think of $X_i$ as the set of possible actions that player $i$ may choose, and the set of outcomes is the product
\[
X=\prod_{i\in I}X_i,
\]
where $X$ has the product topology. We also define player $i$'s utility over outcomes to be a continuous function $u_i\colon X\longrightarrow\mathbb R$. We assume that each $X_i$ (and hence the product) is compact and connected. As is standard, we denote by $-i$ ``all players other than $i$,'' so 
\[
X_{-i}=\mathop{\prod_{j\in I}}_{j\not=i}X_j
\]
is the set of possible action profiles that all players other can $i$ can select, and $x_{-i}$ will denote any point in $X_{-i}$. Set-theoretic concerns may arise when $I$ is infinite, but in this paper, $I$ is always finite.

To each point $x_{-i}\in X_{-i}$, we associate player $i$'s \textit{best-response set}
\[
\BR_i(x_{-i})=\{x_i\in X_i\colon u_i(x_i,x_{-i})=\sup_{y\in X_i}u_i(y, x_{-i})\}.
\]
Informally, the best-response set is the set of $x_i$ values that result in player $i$ receiving the greatest possible payoff conditional on all other players choosing $x_{-i}$.\footnote{For any $x_{-i}$, the best-response set is nonempty if $X_i$ is compact.} (``The {best responses} to $x_{-i}$.'') In general, a player can have multiple best responses, so $\BR_i$ may not be a function. We refer to the collection of all $\BR_i(x_{-i})$ as player $i$'s \textit{best-response correspondence}. See section~3 for more discussion of this point. A point $x\in X$ is a \textit{pure-strategy Nash equilibrium} if for any $i$,
\[
x_i\in\BR_i(x_{-i}),
\]
where $x_i$ is the $i$th coordinate of $x$ and $x_{-i}$ is $x$ with the $i$th coordinate removed. Intuitively, $x$ is a Nash equilibrium if each player is best-responding to all other players. As mentioned previously, the Kakutani and Fan-Glicksberg fixed-point theorems are standard techniques to show Nash equilibrium existence in games that satisfy certain convexity assumptions.

The difficulty with nonconvex strategy spaces is that players' best responses may get ``twisted'' such that they fail to intersect. For example, consider a game where each player's strategy set is the unit circle $S^1\subset\mathbb R^2$, and players' utility functions are given by
\begin{align*}
u_1&=-||x_1-x_2||&
u_2&=-||x_1+x_2||.
\end{align*}
Here $||.||$ denotes the Euclidean norm, so player~1 wants to match $x_2$, while player~2 wants to pick the antipodal point from $x_1$. The two best-response functions that arise from $u_1$ and $u_2$ are
\begin{align*}
\BR_1(x)&=x&\BR_2(x)&=-x,
\end{align*}
and if we embed their graphs in the 2-torus such that coordinate $i$ serves as the domain of $\BR_i$, then any pure-strategy Nash equilibria are exactly the intersections of both graphs. Each graph is an embedding of $S^1$, and if $\phi_i\colon S^1\longrightarrow T^2$ is the embedding corresponding to $\BR_i$, then
\begin{align*}
\phi_1(x)&=(x,x)&
\phi_2(x)&=(x,-x).
\end{align*}
However, $0\not\in S^1$, so $\phi_1$ never equals $\phi_2$, and the game does not have a pure-strategy Nash equilibrium. Graphically, both embeddings lie opposite each other on the torus, so they do not intersect.

Algebraic topology provides tools that can formalize this intuition. Two continuous maps $f,g\colon X\longrightarrow Y$ are \textit{homotopic} if there exists a continuous map $H\colon X\times[0,1]\longrightarrow Y$ such that $H(x,0)=f(x)$ and $H(x,1)=g(x)$. The function $H$ is a \textit{homotopy} or \textit{homotopy equivalence} between $f$ and $g$. If $f$ is homotopic to a constant map, then we say $f$ is \textit{null-homotopic}, and when the identity map on a space $X$ is null-homotopic, we say that $X$ is \textit{contractible}. Intuitively, a homotopy is a continuous deformation, and homotopy equivalence is useful for our purposes because it preserves the amount of ``twistedness'' in $f$ and $g$. My approach to proving equilibrium existence is to show that (1) some well-chosen collection of best-response correspondences guarantees a pure-strategy Nash equilibrium (2) for any game with homotopic best-response correspondences (3) even in situations with unusual strategy sets.

As mentioned previously, my main technique is Nielsen theory, which establishes, for sufficiently nice spaces $X$ and for any continuous self-map $f\colon X\longrightarrow X$, a non-negative integer $N(f)$ such that any function homotopic to $f$ has at least $N(f)$ fixed points. In this context, ``sufficiently nice space'' means a compact, connected Euclidean neighborhood retract (\enr), as I discuss in the next paragraph. See \textcite{jezierski-marzantowicz-2006} for more information on Nielsen theory and the topology of fixed points. The main result that we use from Nielsen theory is the following.

\begin{lemma}
When $f$ is constant, $N(f)=1$. In other words, if Nielsen theory applies on $X$ and if $g\colon X\longrightarrow X$ is null-homotopic, then $g$ has a fixed point.
\end{lemma}

\noindent See \textcite[][p.~120]{jezierski-marzantowicz-2006} for a statement of this lemma and related theory. The result allows us to generalize the Brower and Kakutani fixed-point theorems to allow for nonconvexities because it means that null-homotopic functions can never ``twist'' around a space in a way that wrecks our Nash equilibrium. 

An \enr\ is a closed set $X\subset\mathbb R^n$ such that there exists an open set $U\supset X$ and a continuous map $r\colon U\longrightarrow X$ satisfying $r|_X=\id_X$, where $\id_X$ is the identity map on $X$. In other words, $X$ is a retract of some open subset of Euclidean space. The class of \enr s is large: examples include all manifolds and any finite CW complex. We make use of the following result.

\begin{lemma}
A finite product of \enr s is \enr.
\end{lemma}

\begin{proof}
Let $X\subset\mathbb R^{m}$ and $Y\subset\mathbb R^{n}$ be \enr s, and let $U$ and $V$ be open sets containing them with continuous maps $r$ and $s$ taking $U,V$ to $X,Y$ respectively. Then $X\times Y$ is closed in $\mathbb R^{m+n}$, and $U\times V$ is open in $\mathbb R^{m+n}$. From continuity of $r$ and $s$, the map $r\times s\colon U\times V\longrightarrow X\times Y$ is continuous, and $(r\times s)|_{X\times Y}=r|_X\times s|_Y=\id_X\times \id_Y=\id_{X\times Y}$. This means that $X\times Y$ is an \enr, and the general case follows by induction.
\end{proof}

\begin{corollary}
A finite product of compact, connected \enr s is a compact, connected \enr.
\end{corollary}

\section{An Example Theorem}

\noindent I begin with an example case to illustrate the mathematical intuition before handling the general case in Section~3. The result follows easily from the assumptions of the theorem, Lemma~1, and Corollary~3.

\begin{theorem}
Suppose that $(X_i)$ is a finite collection of compact, connected \enr s, and suppose we have continuous utility functions $u_i\colon X\longrightarrow\mathbb R$ satisfying the following property: for each $i$, we can write player $i$'s best-response correspondence as a continuous function $\BR_i\colon X_{-i}\longrightarrow X_i$. Then if all $\BR_i$ functions are null-homotopic, the game has a pure-strategy Nash equilibrium.
\end{theorem}

\begin{proof}
From Corollary~3, we know that $X=X_1\times X_2\times\dots\times X_n$ is a compact, connected \enr, so Nielsen theory applies to self-maps defined on $X$. Consider the function $\BR_i$. This function has domain $X_{-i}$, and we extend it to $X$ via
\[
\tilde \BR_i(x_i,x_{-i})=\BR_i(x_{-i}).
\]
We know there exists a homotopy $H_i\colon X_{-i}\times[0,1]\longrightarrow X_i$ between $\BR_i$ and a point, and we extend it to a homotopy $\tilde H_i\colon X\times[0,1]\longrightarrow X_i$ in the same way:
\[
\tilde H_i(x_i,x_{-i},t)=H_i(x_{-i},t).
\]
Then consider the functions
\begin{align*}
\phi&=\tilde\BR_1\times\tilde\BR_2\times\dots\times\tilde\BR_n&
H&=\tilde H_1\times\tilde H_2\times\dots\times\tilde H_n.
\end{align*}
The function $\phi$ maps $X$ to itself and is homotopic to a constant map via $H$. From Lemma~1, $\phi$ must have a fixed point, and any fixed point of $\phi$ is a pure-strategy Nash equilibrium.
\end{proof}

\section{The General Case}

In the general case, the best-response correspondence does not need to be a function. Nevertheless, with some minor adjustments, we can adapt the approach of the previous section to this case. See \textcite{gorniewicz-2006} for more information on applying Nielsen theory to correspondences. For a topological space $Y$, let $C(Y)$ be the set of nonempty, closed subsets of $Y$ equipped with the upper Vietoris topology, where we take as a basis the sets $V\subset C(Y)$ such that there exists an open set $U\subset Y$ with $V=\{K\in C(Y)\colon K\subset U\}$. If $Y$ is compact Hausdorff, then $C(Y)$ contains exactly the compact, nonempty subsets of $Y$. A \textit{correspondence} is a function $f\colon X\longrightarrow C(Y)$, and if $f$ is continuous, our definition coincides with the traditional economics notion of a compact-valued, upper-hemicontinuous correspondence. We say that a correspondence $f$ is an \textit{m-map} if (1) it is continuous and (2) for every output $f(x)\subset Y$, there exists an open set $U\supset f(x)$ such that every loop in $U$ is null-homotopic in $Y$ under a homotopy that fixes endpoints.\footnote{Formally, a loop in $Y$ is a continuous function $l:[0,1]\longrightarrow Y$ with $l(0)=l(1)$. If $H$ is the homotopy, fixing endpoints means that $H(0,t)=l(0)$ and $H(1,t)=l(1)$. Fixing endpoints is important because every loop is homotopic to a point under a homotopy that does not fix endpoints, e.g.\ $H(x,t)=l(tx)$.} For context, this property implies that all output sets $f(x)$ are semi-locally simply connected.

A set $X$ is \textit{acyclic} if its homology groups are equal to that of a point: $H_0(X)=\mathbb Z$ and for $n\geq 1$, $H_n(X)=0$. Any contractible (and therefore any convex) set is acyclic. We need to prove two lemmas before we can establish the general result.

\begin{lemma}
A finite product of $m$-maps is an $m$-map.
\end{lemma}

\begin{proof}
Let $f\colon X\longrightarrow C(Y)$ and $g\colon W\longrightarrow C(Z)$ be two $m$-maps. Continuity of $f$ and $g$ implies continuity of $f\times g$. Consider $(x,w)\in X\times W$, and suppose that $U$ and $V$ are the open sets around $f(x)$ and $g(w)$ given by the definition of an $m$-map. If $\phi\colon S^1\longrightarrow U\times V$ is a loop in $Y\times Z$, then we can write $\phi=(\phi_U,\phi_V)$, where $\phi_U$ and $\phi_V$ are loops in $U$ and $V$ respectively. If $H_1$ and $H_2$ are homotopies between these two loops and two constant functions, then the product $H_1\times H_2$ is a homotopy between $\phi$ and a constant. This means $f\times g$ is an $m$-map, and the general case follows by induction.
\end{proof}

\begin{lemma}
A finite product of acyclic sets is acyclic.
\end{lemma}

\begin{proof}
Let $X$ and $Y$ be acyclic. Then $H_n(X)$ and $H_n(Y)$ are either 0 or $\mathbb Z$, so $\tor_1(H_m(X),\penalty0 H_n(Y))=0$. Applying the K\"unneth formula for singular homology gives us an exact sequence
\[
0\longrightarrow\prod_{i+j=n}H_i(X)\otimes H_j(Y)\longrightarrow H_n(X\times Y)\longrightarrow 0
\]
The tensor product will always be 0 unless both $H_i(X)$ and $H_j(Y)$ are $\mathbb Z$, and therefore the direct product is $\mathbb Z$ if $i=j=n=0$ and 0 otherwise. By exactness, the middle arrow is an isomorphism, so $X\times Y$ is acyclic. The general case follows by induction.
\end{proof}

When applying Nielsen theory to a correspondence $f\colon X\longrightarrow C(X)$, we require $f$ to be a compact-valued, acyclic-valued $m$-map, and any homotopy $H\colon X\times[0,1]\longrightarrow C(X)$ must be a compact-valued, acyclic-valued $m$-map as well. Further, a fixed point in this case is a point $x$ such that $x\in f(x)$, and $N(f)$ is a lower bound on the number of points $x$ with $x\in g(x)$, where $g$ is any compact-valued, acyclic-valued $m$-map that is homotopic to $f$ as described. Notably, if $f$ is a constant correspondence, then $N(f)$ is still 1. See \textcite[][pp.~182, 183]{gorniewicz-2006} for details. At this point, we are ready to state and prove the main theorem of the paper.

\begin{theorem}
Suppose that $(X_i)$ is a finite collection of compact, connected \enr s, and suppose we have continuous utility functions $u_i\colon X\longrightarrow\mathbb R$ satisfying the following property: for each $i$, player $i$'s best response is an acyclic-valued $m$-map $\BR_i\colon X_{-i}\longrightarrow C(X_i)$. Then if all $\BR_i$ correspondences are null-homotopic under an acyclic-valued $m$-map homotopy, the game has a pure-strategy Nash equilibrium.
\end{theorem}

\begin{proof}
The proof is exactly the same as that of Theorem~4, except that Lemmas~5 and 6 guarantee that $\phi$ and $H$ will be acyclic-valued $m$-maps. Because the $X_i$ are compact Hausdorff, each $C(X_i)$ contains only nonempty compact sets, so $\BR_i$ and the associated homotopies will be compact-valued. Both $\phi$ and $H$ will be compact-valued because any product of compact sets is compact.
\end{proof}

Theorem~7 is very general. Compact, connected \enr s often coincide with what we think of as ``nice'' subsets of Euclidean space, and being acyclic-valued means that the best-response correspondence values should resemble points topologically. For example, the best-response correspondence can never output a circle or sphere. The homotopy criterion is perhaps the least straightforward, and intuitively, it means that the best responses should not lead players to perpetually ``chase'' each other around the outcome space. In our example involving $S^1\times S^1$ earlier, Theorem~7 does not apply because the $\phi_1$ and $\phi_2$ embeddings ``twist'' around the torus to avoid each other in a way that is possible only when we cannot deform the best-response curves $\BR_1$ and $\BR_2$ into a single point. Corollary~8 captures the same ideas and may involve more familiar concepts for some readers.

\begin{corollary}
Suppose that $(X_i)$ is a finite collection of compact and connected manifolds, and suppose we have continuous utility functions $u_i\colon X\longrightarrow\mathbb R$ satisfying the following properties for every player $i$:
  \begin{enumerate}[nosep, label=(\arabic*)]
  \item Player $i$'s best response is a continuous function $\BR_i:X_{-i}\longrightarrow C(X_i)$;
  \item Every output value $\BR_i(x_{-i})$ is contractible and has a contractible neighborhood in $X_i$; and
  \item There exists a homotopy that takes $\BR_i$ to a constant map and whose output values---which are elements of $C(X_i)$---are all contractible and have a contractible neighborhood in $X_i$.
  \end{enumerate}
Then the game has a pure-strategy Nash equilibrium.
\end{corollary}

\section{An Application}

In addition to showing the existence of Nash equilibrium, Theorem~7 (and Corollary~8) can be used to show that certain maps are not null-homotopic. If a game does not have a pure-strategy Nash equilibrium, then at least one best-response correspondence must not be homotopic to a point, and I use this idea to prove the fundamental theorem of algebra. The proof is similar to standard topological proofs, although it does not require computing $\pi_1(S^1)$.

\begin{theorem}
Let $X$ be a compact, connected \enr. If there exists a continuous self-map $f\colon X\longrightarrow X$ without a fixed point, then $X$ is not contractible.
\end{theorem}

\begin{proof}
Consider a game with two copies of $X$ as strategy spaces and utility functions satisfying
\begin{align*}
u_1(x_1,x_2)&=-||x_1-x_2|| & u_2(x_1,x_2)&=-||f(x_1)-x_2||.
\end{align*}
Then the best-response functions are given by $\BR_1=\id_X$ and $\BR_2=f$. The game has no pure-strategy Nash equilibrium because any pure-strategy equilibrium is a fixed point of $f$. However, if $X$ is contractible, then $\BR_1$ and $\BR_2=\id_X\circ f$ are both null-homotopic, which by Theorem~7 would guarantee a pure-strategy Nash equilibrium for the game. By contrapositive, $\id_X$ must not be null-homotopic.
\end{proof}

\begin{corollary}
The $n$-sphere is not contractible.
\end{corollary}

\begin{proof}
The antipodal map has no fixed points.
\end{proof}

Corollary~10 is the result we need to prove the fundamental theorem of algebra. We show that if a polynomial with complex coefficients has no root, it would mean that the unit circle is contractible.

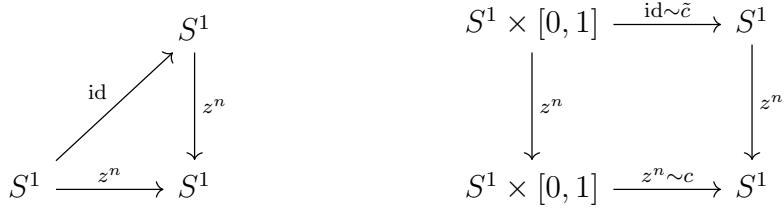
\begin{figure}[t]
\centerline{\bfseries Figure 1: A Covering Map on the Circle\strut}
\medskip
\noindent\hfil\begin{tikzcd}[sep=3.5em]
& S^1\arrow[d,"z^n"]\\
S^1\arrow[ur, "\id"]\arrow[r,"z^n"]& S^1
\end{tikzcd}%
\hfil
\begin{tikzcd}[sep=3.5em]
S^1\times[0,1]\arrow[r, "\id\sim\tilde c"]\arrow[d, "z^n"] & S^1\arrow[d, "z^n"]\\
S^1\times[0,1]\arrow[r, "z^n\sim c"]& S^1
\end{tikzcd}
\hfil
\caption{The function $z^n$ is a covering map of the unit circle. \textbf{Left:} The map $z^n$ lifts to the identity. \textbf{Right:} The homotopy between $z^n$ and the constant map $c$ lifts to a homotopy between the identity map and the constant map $\tilde c$. Such a homotopy would mean the unit circle is contractible, which contradicts Corollary~10.}
\end{figure}

\begin{theorem}[Fundamental Theorem of Algebra]
Every nonconstant polynomial with coefficients in $\mathbb C$ has a root in $\mathbb C$.
\end{theorem}

\begin{proof}
For the sake of contradiction, let $P(z)$ be an $n$th-degree complex polynomial without a root. Then for $z\in S^1$, $f(z)=P(z)/|P(z)|$ maps the unit circle to the unit circle. The function $H_1(z,t)=f(tz)$ is a homotopy between $f$ and $P(0)/|P(0)|$, so $f$ must be null-homotopic. At the same time,
\[
H_2(z,t)=f\bigg(\frac zt\bigg)=\frac{t^nP(z/t)}{|t^nP(z/t)|}
\]
is a homotopy between $f$ and $z^n$, which means $z^n$ is null-homotopic. However, as shown in Figure~1, $z^n$ is a covering map on $S^1$, so the homotopy $z^n\sim c$, where $c$ is a constant map, lifts to a homotopy $\id\sim \tilde c$, where $\tilde c$ maps $S^1$ into the preimage $(z^n)^{-1}c(S^1)$. The preimage under $z^n$ of a singleton is a discrete set of $n$ points, and because $S^1$ is connected, the image of $\tilde c$ must be a single point, which means $\tilde c$ is also a constant map. It follows that the induced homotopy contradicts Corollary~10, so $P$ must have a root.
\end{proof}

\section{Conclusion}

We have established that pure-strategy Nash equilibrium exists for a large class of games with nonconvexities, specifically those where the best-response correspondences do not get ``twisted.'' Because any convex subset of $\mathbb R^n$ is contractable and therefore acyclic, the result in Theorem~7 is strictly more general than results obtained using the Kakutani fixed-point theorem. One possible direction for future research is to expand the results here by allowing maps from different homotopy equivalence classes with nonzero Nielsen number. Because calculating Nielsen numbers is difficult, this is potentially a complicated task. Leaving behind convexity requirements opens the door to unexpected outcomes, but Theorem~7 suggests that we do not need to worry for sufficiently well-behaved games.

\section*{References}

{
\def\mkbibnamefamily#1{\textsc{#1}}
\def\mkbibnamegiven#1{\textsc{#1}}
\def\mkbibnameprefix#1{\textsc{#1}}
\def\mkbibnamesuffix#1{\textsc{#1}}

\printbibliography[heading=none]

\vskip-\lastskip

}

\end{document}